\documentclass[preprint,12pt]{elsarticle}

\usepackage{graphics}
\usepackage{graphicx}
\usepackage{epsfig}
\usepackage[dvipsnames]{xcolor}
\usepackage{amssymb}
\usepackage{amsthm}
\usepackage{amsmath}
\usepackage{url}

\newcounter{bla}

\journal{Computer Physics Communications}

\begin{document}

\begin{frontmatter}



\title{A general recursion for integrals involving products of Hermite polynomials and its applications}

\author[a,b,c]{Tran Duong Anh-Tai\corref{author1}}
\cortext[author1]{\textit{E-mail address:} tai.d.tran-1@ou.edu}
\author[d,e]{Phan Quang Son}
\author[f]{Le Minh Khang}
\author[g,h]{Nguyen Duy Vy\corref{author2}}
\cortext[author2]{\textit{E-mail address:} nguyenduyvy@vlu.edu.vn}
\author[i,j]{Vinh N.T. Pham\corref{author3}}
\cortext[author3]{Corresponding author.\textit{E-mail address:} vinhpnt@hcmue.edu.vn}
\address[a]{Quantum Systems Unit, OIST Graduate University, Onna, Okinawa 904-0495, Japan}
\address[b]{Center for Quantum Research and Technology, The University of Oklahoma, 440 W. Brooks Street, Norman, Oklahoma 73019, USA}
\address[c]{Homer L. Dodge Department of Physics and Astronomy, The University of Oklahoma, 440 W. Brooks Street, Norman, Oklahoma 73019, USA}
\address[d]{Department of Theoretical Physics, Faculty of Physics and Engineering Physics, University of Science, Ho Chi Minh City 700000, Vietnam}
\address[e]{Vietnam National University, Ho Chi Minh City 700000, Vietnam}
\address[f]{OIST Graduate University, Onna, Okinawa 904-0495, Japan}
\address[g]{Laboratory of Applied Physics, Science and Technology Advanced Institute, Van Lang University, Ho Chi Minh City, Vietnam}
\address[h]{Faculty of Applied Technology, Van Lang School of Technology, Van Lang University, Ho Chi Minh City, Vietnam}
\address[i]{Postgraduate Studies Office, Ho Chi Minh City University of Education, Ho Chi Minh City, Vietnam}
\address[j]{Department of Physics, Ho Chi Minh City University of Education, Ho Chi Minh City, Vietnam}

\begin{abstract}
This study presents the derivation of a recursive formula for integrals of products of $N$ Hermite polynomials, establishing a numerically stable scheme for their accurate evaluation in computer codes. The derivation is notably simple and leverages solely the well-established properties of Hermite polynomials and the method of integration by parts. Importantly, our formulation completely circumvents explicit factorials, thereby preventing potential numerical instabilities and overflows, while facilitating high-precision computations for large indices. These findings are of significant relevance to a variety of areas in physics and mathematics. In particular, they offer an efficient and accurate framework for calculating two- and three-body matrix elements in ab initio simulations of few-body systems under a 1D harmonic confinement using the Configuration Interactions approach. A numerical subroutine implementing the recursive formula is provided as supplemental material.
\end{abstract}
\end{frontmatter}

\section{Introduction}\label{sec1}

The study of correlated few-body systems in one-dimensional (1D) space from first principles has received great attention due to the remarkable quantum phenomena occurring from particle-particle correlations. Moreover, the physics of 1D systems differs significantly from that of 3D systems \cite{giamarchi2003quantum}. These systems have the advantage of being experimentally realizable utilizing ultra-cold atomic setups \cite{serwane2011deterministic,wenz2013few,zurn2013pairing,zurn2012fermionization,murmann2015antiferromagnetic,murmann2015two,bergschneider2019experimental}. For a more comprehensive review, we refer the reader to Refs.~\cite{sowinski2019one,mistakidis2023few}. This study concentrates on ab initio simulations of 1D systems comprising a few interacting particles confined in harmonic traps, utilizing the Configuration Interactions approach \cite{anhtai2023quantum,Tai2024inprepPRX,tran2025quantum,rojo2024few,garcia2014quantum,rojo2020static,garcia2013quantum,garcia2013sharp,garcia2014distinguishability,garcia2018relaxation,garcia2016non,tran2025quantum}. A crucial aspect of this approach entails evaluating integrals comprising products of Hermite polynomials, with a specific focus on three particular integrals as follows

\begin{align}
    \label{eq:desired_integral}
    W_{ijk\ell} =& A_{ijk\ell} \int\limits_{-\infty}^{+\infty} e^{-2x^2} H_{i}(x)H_{j}(x)H_{k}(x)H_{\ell}(x) dx, \\ 
    \label{eq:desired_integral_6H}
    U_{ijk\ell mn} =&B_{ijk\ell mn} \int\limits_{-\infty}^{+\infty} e^{-3x^2} H_{i}(x)H_{j}(x)H_{k}(x)H_{\ell}(x)H_{m}(x)H_{n}(x) dx,
\end{align}
and 
\begin{align}
\label{eq:desired_integralp}
    Y_{ijk\ell} = A_{ijk\ell} \int\limits_{-\infty}^{+\infty} e^{-2x^2} \left[H^\prime_{i}(x)H_{j}(x) - H_{i}(x)H^\prime_{j}(x) \right] \left[H^\prime_{k}(x)H_{\ell}(x) - H_{k}(x)H^\prime_{\ell}(x) \right] dx.
\end{align}
Here $A_{ijk\ell} = \dfrac{1}{\pi\sqrt{2^{i+j+k+\ell}i!j!k!\ell !}}$ and ${B_{ijk\ell mn} = \dfrac{1}{\pi\sqrt{\pi}\sqrt{2^{i+j+k+\ell+m+n}i!j!k!\ell ! m! n!}}}$ are the normalization coefficients, $H_i(x)$ is the physicists' Hermite polynomial of degree $i$ and $H^\prime_i(x)$ denotes the first-order derivative of $H_i(x)$. Physically, $W_{ijk\ell}$, and $Y_{ijk\ell}$ respectively constitute the matrix elements of the s- and p-wave two-body contact interaction potentials in bosonic~\cite{olshanii1998atomic}, and fermionic~\cite{kanjilal2004nondivergent} systems under a one-dimensional harmonic confinement. Analogously, $U_{ijk\ell mn}$ accounts for the matrix elements of three-body interactions, which are particularly relevant in the study of anyonic systems~\cite{harshman2020anyons}. The indices $i,j,k,\ell,m,n$ are non-negative integers of 0, 1, \dots, $M$, where $M$ is the largest degree of the Hermite polynomial which can be very large to achieve the desired numerical convergence in practice. While the integrals $W_{ijk\ell}$, $Y_{ijk\ell}$ and $U_{ijk\ell mn}$ can, in principle, be computed numerically using efficient quadrature methods available in various computational packages and software, such brute-force methods are highly sensitive to the choice of integration schemes, the number of grid points, and the integration interval. Symbolic math toolboxes might offer an alternative, but for large values of $M$, this approach becomes impractically slow, as the polynomially growing number of distinct integrals requires a substantial amount of computing time. 

In recent years, substantial effort has been dedicated to deriving analytical formulas for the integral $W_{ijk\ell}$. Early approaches concentrated on deriving analytical formulas for a single set of indices $(i,j,k,\ell)$, as exemplified by the foundational studies of Talmi, Brody, and Moshinsky \cite{talmi1952nuclear,moshinsky1959transformation,brody1965tables}, which are today named as the Talmi-Brody-Moshinsky expansion. A variation of this expansion is presented in Ref.~\cite{rammelmuller2023modular}. Moreover, Wang et al.~\cite{wang2009integrals} employed generating functions for the products of two Hermite polynomials to derive analytical expressions for $W_{ijk\ell}$. An analytical expression for the integral $W_{ijk\ell}$ was recently derived for a single set of indices $(i,j,k,\ell)$ in Ref.~\cite{rojo2020static}. This analytical expression is based on the established solution for the integral of a product of three Hermite polynomials~\cite{titchmarsh1948some} and relies on the logarithm of the gamma function and factorials to mitigate numerical overflows associated with factorials of large integers. Although this result is fairly useful, its practical implementation is hindered by its considerable algebraic complexity. In particular, the appearance of additional phase factors when evaluating the logarithm of the gamma function for negative arguments significantly complicates efficient implementation in computer codes. Consequently, due to the explicit use of integer factorials in these established analytical formulas, numerical issues emerge as the degree of the Hermite polynomials becomes large. While techniques like Stirling’s approximation or the use of logarithms of gamma functions might partially mitigate these challenges to some extent, the explicit appearance of factorials in the analytical expressions remains problematic, particularly at high polynomial degrees. Consequently, it is preferable to avoid the appearance of factorials whenever possible, particularly in numerical calculations. To our knowledge, no analytical expressions for $Y_{ijk\ell}$ or $U_{ijk\ell mn}$ are present in the existing literature, and the methodology outlined in Ref.~\cite{rojo2020static} does not easily adapt to the integrals $Y_{ijk\ell}$ and $U_{ijk\ell mn}$. 

Apparently, there is a clear need for the development of an efficient, robust, and numerically stable approach for computing $W_{ijk\ell}$, $Y_{ijk\ell}$ and $U_{ijk\ell mn}$, applicable not just in quantum physics but also in related computational contexts more broadly. Such an approach must maintain high numerical precision at large Hermite polynomial degrees while circumventing the intrinsic limits of symbolic or brute-force integration and the numerical instability arising from large-integer factorials.  

To address the aforementioned challenges, we formulate a general recursive relation for integrals involving products of $N$ Hermite polynomials, exploiting their recurrence properties and the integration-by-parts technique in this present work. The key advantage of our formulation is that the recursive formula entirely eliminates the explicit use of factorials, thus effectively avoiding numerical instabilities and overflows associated with factorials of large integers. Notably, the first integral in the recursion manifests as a simple Gaussian integral with a closed analytical expression, from which all subsequent non-zero integrals to be obtained recursively, rendering the numerical evaluation completely independent of any quadrature methods. In addition, the symmetry of the integration interval induces a selection rule, identifying only the non-zero integrals within the recursion, allowing for an efficient implementation in computer codes. Having the general recursion for the integral of products of $N$ Hermite polynomials, the integrals $W_{ijk\ell}$ and $U_{ijk\ell mn}$ corresponds to the cases $N=4$ and $N=6$, respectively. Meanwhile, $Y_{ijk\ell}$ can be expressed in terms of $W_{ijk\ell}$ and hence evaluated via its recursion. Finally, to facilitate practical usage, we provide a Mathematica notebook and a Python script implementing the recursive formulas, available at \url{https://github.com/Tran-Duong-Anh-Tai/integral_Hermite_polynomials}.

\section{Results and Discussion}\label{sec2}

In this section, we rigorously present the derivation of the recursive formula for the generalized integral of products of $N$ Hermite polynomials 
\begin{equation} 
    \label{eq:generalized_integral}
    T_{a_1\dots a_N} = A_{a_1\dots a_N} \displaystyle\int\limits_{-\infty}^{+\infty} \exp \left(-\dfrac{N}{2} x^2 \right) \prod\limits_{i=1}^N H_{a_i}(x) dx,
\end{equation}
where 
\begin{equation}
    A_{a_1\dots a_N} = \dfrac{1}{\sqrt{\sqrt{\pi^N}} \sqrt{2^{\sum\limits_{i=1}^N a_i}  \prod\limits_{i=1}^N (a_i!)}},
\end{equation}
is the normalization coefficient, and $H_{n}(x)$ denotes the physicists' Hermite polynomial of degree $n$. Note that there are $N\geq 2$ subscripts ($a_1,\dots,a_N$) in the notations $A_{a_1\dots a_N}$ and $T_{a_1\dots a_N}$. Also we will put the subscripts in a pair of parentheses as needed to prevent ambiguity in the following discussion (i.e. $A_{(a_1-1)a_2\dots (a_{i}-1)\dots a_N}$). 

With the use of the relation \cite{weber2003essential}
\begin{equation}
    H_n(x) = 2xH_{n-1}(x) - 2(n-1)H_{n-2}(x),
\end{equation}
the integral~\eqref{eq:generalized_integral} can be rewritten as

\begin{align} 
 	\label{eq:T}
 	T_{a_1\dots a_N} = & \underbrace{2{A_{a_1\dots a_N}}\int\limits_{-\infty}^{+\infty} x \exp \left(-\dfrac{N}{2} x^2 \right)  H_{a_1-1}(x) \prod\limits_{i=2}^N H_{a_i}(x) dx}_{L} \nonumber \\ & 	- 2(a_1 -1){A_{a_1\dots a_N}}\int\limits_{-\infty}^{+\infty} \exp \left(-\dfrac{N}{2} x^2 \right) H_{a_1-2}(x)\prod\limits_{i=2}^N H_{a_i}(x) dx.
\end{align}
Let us integrate $L$ by parts  
\begin{align}
	L  = & \underbrace{-\dfrac{2}{N} A_{a_1\dots a_N} \exp \left(-\dfrac{N}{2} x^2 \right) H_{a_1-1}(x)   \prod\limits_{i=2}^N H_{a_i}(x) \bigg\rvert_{-\infty}^{+\infty}}_{0} \nonumber \\ &
	+\dfrac{2}{N} A_{a_1\dots a_N} \displaystyle\int\limits_{-\infty}^{+\infty} \exp\left(-\dfrac{N}{2} x^2 \right) \dfrac{d}{dx} \left[ H_{a_1-1}(x) \prod\limits_{i=2}^N H_{a_i}(x) \right ] dx,
\end{align} 
with the use of the relation
\begin{equation}
    \dfrac{d}{dx} \exp \left(-\dfrac{N}{2} x^2 \right) = -Nx \exp \left(-\dfrac{N}{2} x^2 \right).
\end{equation}
Using of the property \cite{weber2003essential}
\begin{equation}
    \label{eq:derivativeH}
    \dfrac{d}{dx} H_n(x) = 2nH_{n-1}(x),
\end{equation}
yields
\begin{align}
    \label{eq:simplifiedL}
    L = \dfrac{4}{N} A_{a_1\ldots a_N}\int_{-\infty}^{\infty} \exp\left({-\frac{N}{2}x^2}\right) \sum_{j=1}^N
    \Bigg[(a_j - \delta_{j1}) \prod_{i=1}^N H_{a_i - \delta_{i1} - \delta_{ij}}(x) \Bigg] dx .
\end{align}
Substituting Eq.~\eqref{eq:simplifiedL} into Eq.~\eqref{eq:T}, we obtain
\begin{align}
    \label{eq:TT}
    T_{a_1\ldots a_N}=&\; 2(a_1 - 1)\!\left(\frac{2}{N} - 1\right)A_{a_1\ldots a_N}\int_{-\infty}^{+\infty}\exp\left({-\frac{N}{2}x^2}\right)H_{a_1 - 2}(x)\prod_{i=2}^{N} H_{a_i}(x) dx  \nonumber \\&  +\frac{4}{N} A_{a_1\ldots a_N}\int_{-\infty}^{+\infty} \exp\left({-\dfrac{N}{2}x^2}\right) \sum_{j=2}^{N} \Bigg[ a_j \prod_{i=1}^N H_{a_i - \delta_{i1} - \delta_{ij}}(x)\Bigg] dx.
\end{align}
We note that the coefficient appearing in front of each integral in Eq.~\eqref{eq:TT} admits a recursive representation, namely,
\begin{align}
2(a_1 -1)A_{a_1\dots a_N} & = \dfrac{2(a_1 -1)}{\sqrt{\sqrt{\pi^N}}  \sqrt{2^2 2^{(a_1-2) + a_2 \dots + a_N} \prod\limits_{i=1}^N (a_i!)}} \nonumber \\ &  = \sqrt{\dfrac{a_1-1}{a_1}} A_{(a_1-2)a_2\dots a_N} \label{eq:coeff1} \\ 
4a_iA_{a_1\dots a_N} & =  2 \sqrt{\dfrac{a_i}{a_1}} A_{(a_1-1)a_2\dots (a_{i}-1)\dots a_N}\label{eq:coeff2}
\end{align}
Substituting Eqs.\eqref{eq:coeff1} and \eqref{eq:coeff2} into Eq.~\eqref{eq:TT}, we obtain the recursive formula for the integral of product of $N$ Hermite polynomials $T_{a_1\dots a_N}$ as follows:
\begin{align}
    \label{eq:recursiveT}
	T_{a_1\dots a_N} =  \left(\dfrac{2}{N}-1 \right) \sqrt{\dfrac{a_1-1}{a_1}} T_{(a_1-2)a_2\dots a_N} + \dfrac{2}{N}\sum_{i=2}^{N} \sqrt{\dfrac{a_i}{a_1}} T_{(a_1-1)\dots(a_i-1)\dots a_N}. 
\end{align}
It is noted that, during the recursion, any term generated by the recurrence that contains a negative index $(a_i<0)$ is defined to vanish. This convention is consistent with the Hermite polynomial recurrence relations and ensures that all recursive expressions are well defined~\cite{weber2003essential}. Furthermore,  since the integral $T_{a_1\dots a_N}$ is invariant under a simultaneous permutation of its indices, sorting them into ascending or descending order enables the efficient evaluation of only unique nonzero integrals. Importantly, since the recursive formula in Eq.~\eqref{eq:recursiveT} relies solely on elementary mathematical operations and avoids the explicit use of factorials, it completely circumvents the numerical overflow associated with factorials of large integers.

\begin{table}[htbp]
	\centering
	\caption{\label{tab:table1} The values of the non-zero integrals $W_{ijk\ell}$ for the few levels $V$. For brevity, the indices are normally sorted as $0\leq\ell\leq k\leq j\leq i$. Note that the integrals with the indices that are the permutations of the ordered set $(i,j,k,\ell)$ are identical. We use symbolic integration in Mathematica (the built-in function Integrate) to evaluate $W_{ijkl}$ for individual sets of indices. The resulting values serve as high-precision references for validating the accuracy of the results obtained from our recursive method. Meanwhile, the Python implementation performs calculations using double-precision floating-point arithmetic. The fifth column reports the absolute difference, $\varepsilon$, between the results obtained using the Python implementation and the corresponding reference values obtained from the symbolic Mathematica calculations. }
    \small
	\begin{tabular}{|c|c|r|r|r|}
		\hline
		\textbf{Level} ${V}$               
		&   ${W_{ijkl}}$
		&   \textbf{Python}
		&  	\textbf{Mathematica}			
		&   $\mathbf{\varepsilon} (\times 10^{-17})$ 
		\\ 
	
		\hline
		   $0$                       
	 	&  $W_{0000}$
		&  $0.3989422804014327 $   
		&  $\frac{1}{\sqrt{2\pi}}$				
		&  $2$							
		\\
		\hline                              
		$2$
		&  $W_{2000}$
		&  $-0.1410473958869391 $ 
		&  $- \frac{1}{4\sqrt{\pi}}$ 				
		&  $2$									
		\\ 
		&  $W_{1100}$
		&  $0.1994711402007163 $ 
		&  $ \frac{1}{2\sqrt{2\pi}} $ 					
		&  $1$	 								
		\\
		
		\hline                             
		$4$
		&  $W_{4000}$   
		&  $0.0610753139878650 $
		&  $\frac{1}{16}\sqrt{\frac{3}{\pi}} $ 		
		&  $2$	 								
		\\ 
		&  $W_{3100}$ 
		&  $-0.1221506279757300 $  
		&  $-\frac{1}{8} \sqrt{\frac{3}{\pi}}$ 		
		&  $5$	 								
		\\ 
		&  $W_{2200}$
		&  $0.1496033551505372 $ 
		&  $\frac{3}{8\sqrt{2\pi}} $ 				
		&  $2$	  								
		\\ 
		&  $W_{2110}$
		&  $0.0705236979434695$ 
		&  $\frac{1}{8\sqrt{\pi}} $ 		
		&  $1$ 								
		\\ 
		&  $W_{1111}$
		&  $0.2992067103010745$ 
		&  $\frac{3}{4\sqrt{2\pi}} $ 				
		&  $4$  								
		\\ 
		
		\hline
		$6$
		&  $W_{6000}$
		&  $ -0.0278769393148870 $ 
		&  $ -\frac{1}{32}\sqrt{\frac{5}{2\pi}} $  	
		&  $2$  									
		\\
		&  $W_{5100}$
		&  $0.0682842769120049 $ 
		&  $ \frac{1}{32} \sqrt{\frac{15}{\pi}} $ 	
		&  $3$								
		\\ 
		&  $W_{4200}$
		&  $ - 0.1079669217097923 $ 
		&  $ -\frac{5}{32} \sqrt{\frac{3}{2\pi}} $ 	
		&  $4$								
		\\ 
		&  $W_{4110}$
		&  $ -0.0916129709817974 $ 
		&  $ - \frac{3}{32} \sqrt{\frac{3}{\pi}} $ 
		&  $3$	
		\\ 
		&  $W_{3300}$
		&  $0.1246694626254477 $ 
		&  $\frac{5}{32} \sqrt{\frac{2}{\pi}} $		
		&  $3$								
		\\ 
		&  $W_{3210}$
		&  $0.0431867686839169 $ 
		&  $\frac{1}{16} \sqrt{\frac{3}{2\pi}} $ 	
		&  $2$   
		\\ 
		&  $W_{3111}$
		&  $-0.0610753139878650  $ 
		&  $-\frac{1}{16} \sqrt{\frac{3}{\pi}} $  	
		&  $2$								
		\\ 
		&  $W_{2220}$
		&  $0.0176309244858673 $ 
		&  $\frac{1}{32\sqrt{\pi}} $ 				
		&  $2$								
		\\ 
		&  $W_{2211}$
		&  $0.1745372476756268 $ 
		&  $\frac{7}{32} \sqrt{\frac{2}{\pi}} $ 		
		&  $3$								
		\\ 
		\hline
		80 &	$W_{(20)(20)(20)(20)}$                       
			&  $0.1227318927419739$   
			&  $0.1227318927419739$					
			&  $0$							
			\\
			\hline
		80 &	$W_{(26)(24)(15)(15)}$                       
			&   $-0.0178894841345128$ 
			&  	$-0.0178894841345128$				
			&  	$0$					
			\\
			\hline                               
        200 &	$W_{(50)(50)(50)(50)}$   
			&  $0.0872886184434822$ 
			&  $0.0872886184434821$ 					
			&  $1$	 								
			\\					
			\hline
        200 &	$W_{(198)(197)(195)(194)}$   
			&  $0.0253004613111027$
			&  $0.0253004613111027$ 					
			&  $0$							
			\\					
			\hline    
		400 &	$W_{(100)(100)(100)(100)}$   
			&   $0.0667942531440904$
			&   $0.0667942531440905$		
			&   $1$
			\\
			\hline  			
        400 &	$W_{(200)(200)(0)(0)}$   
			&   $0.0159055502434488$
			&   $0.0159055502434489$		
			&   $1$
			\\
			\hline  	
		600 &	$W_{(150)(150)(150)(150)}$   
			&  $0.0569408575509921  $
			&  $0.0569408575509922 $		
			&  1 
			\\		
			\hline
        600 &	$W_{(280)(220)(60)(40)}$   
			&  $0.0112661422390370 $
			&  $0.0112661422390371 $		
			&  1
			\\		
			\hline
		800 &	$W_{(200)(200)(200)(200)}$  
			&  $0.0507841297972471 $
			&  $0.0507841297972472	$	
			&  1		\\	
			\hline
        800 &	$W_{(525)(524)(481)(464)}$  
			&  1.312845978231059e-5
			&  1.312845978231137e-5
			&  $0.078$							
			\\	
			\hline
		2000 & $W_{(500)(500)(500)(500)}$  
			&  $0.0350723584392968 $
			&  $0.0350723584392968 $
			&  0					
			\\	
			\hline
        4000 & $W_{(1000)(1000)(1000)(1000)}$  
			&  $0.0263749774869170 $
			&  $0.0263749774869169 $
			&  1 			
			\\	
			\hline 
	\end{tabular}
\end{table} 

It is critical to note that the integral $T_{a_1\dots a_N}$ obeys a selection rule that arises from the symmetry of the integration domain. Since the integral is evaluated over a symmetric interval, only the even integrand yields non-zero results. Consequently, this strictly implies that the product of $N$ Hermite polynomials must be an even function because the Gaussian weight $\exp(-Nx^2/2)$ is inherently even. Using the parity property of Hermite polynomials \cite{weber2003essential}
\begin{equation}
    H_{a_i}(-x) = (-1)^{a_i} H_{a_i}(x),
\end{equation}
one obtains
\begin{equation}
    \prod\limits_{i=1}^N H_{a_i}(-x) = (-1)^{\sum\limits_{i=1}^N a_i} \prod\limits_{i=1}^N H_{a_i}(x) = (-1)^{V} \prod\limits_{i=1}^N H_{a_i}(x),
\end{equation}
where $V=\sum\limits_{i=1}^N a_i$. Consequently, the integrand is even resulting in non-vanishing $T_{a_1\dots a_N}$ if and only if $V$ is an even integer. Hence, $T_{a_1\dots a_N}$ vanishes for odd values of $V$ as the integrand is an odd function. This well-defined parity constraint is preserved by the recursive relation in Eq.~\eqref{eq:recursiveT}, which recursively connects non-zero integrals at level $V$ exclusively to those at level $V-2$. As a result, all non-zero integrals $T_{a_1\dots a_N}$ can be obtained recursively starting from the lowest level $V=0$, for which the integral reduces to a fundamental Gaussian-type integral of the form 
\begin{equation} 
 	T_{0\dots 0} = \dfrac{1}{\sqrt{\sqrt{\pi^N}}} \displaystyle\int\limits_{-\infty}^{+\infty} \exp \left(-\dfrac{N}{2} x^2 \right) dx = \sqrt{\dfrac{2}{N\pi^{N/2-1}}},
\end{equation}
which is known in closed form. Thereby, this enables all subsequent non-zero integrals $T_{a_1\dots a_N}$ to be obtained analytically via the recursion, independent of any specific numerical integration schemes. We have demonstrated that the evaluation of $T_{a_1\dots a_N}$ through this the recursive relation remains free from potential numerical instabilities associated with quadrature methods and large-integer factorials. Furthermore, in the computer science language, the recursive structure facilitates the use of dynamic programming, providing a robust framework for practical implementation of Eq.~\eqref{eq:recursiveT} in computer codes. Overall, the recursive relation in Eq.~\eqref{eq:recursiveT} overcomes the major challenges and ensures that the evaluation of $T_{a_1\dots a_N}$ is not only highly efficient, also computationally stable in practice, which indeed constitutes the central result of the present work. 

Having the general recursive relation for the integral involving the product of $N$ Hermite polynomials, we now turn to three specific cases of interest in the following discussion. In particular, the integral of product of six Hermite polynomials $U_{ijk\ell mn}$ can be accurately computed via the recursive relation

\begin{align}
    \label{eq:recursiveU}
    U_{ijk\ell mn} = \dfrac{1}{3}\left[-2\sqrt{\dfrac{i-1}{i}}U_{(i-2)jk\ell mn} + \sqrt{\dfrac{j}{i}}U_{(i-1)(j-1)k\ell mn} + \sqrt{\dfrac{k}{i}}U_{(i-1)j(k-1)\ell mn}   \right. \nonumber\\ \left. +\sqrt{\dfrac{\ell}{i}}U_{(i-1)jk(\ell-1) mn}  + \sqrt{\dfrac{m}{i}}U_{(i-1)jk\ell(m-1)n}  + \sqrt{\dfrac{n}{i}}U_{(i-1)jk\ell m(n-1)}\right],
\end{align}
meanwhile that of product of four Hermite polynomials $W_{ijk\ell}$ is explicitly given as
\begin{align}
    \label{eq:recursiveW}
    W_{ijk\ell} = \dfrac{1}{2} &\left[- \sqrt{\dfrac{i-1}{i}} W_{(i-2)jk\ell} + \sqrt{\dfrac{j}{i}} W_{(i-1)(j-1)k\ell} \right. \nonumber \\ & \left. + \sqrt{\dfrac{k}{i}} W_{(i-1)j(k-1)\ell}  + \sqrt{\dfrac{\ell}{i}} W_{(i-1)jk(\ell-1)} \right],
\end{align}
with
\begin{align}
    U_{000000} & = \dfrac{1}{\pi\sqrt{\pi}} \int\limits_{-\infty}^{+\infty} \exp(-3x^2) dx = \dfrac{1}{\sqrt{3}\pi},\\
    W_{0000}   & = \dfrac{1}{\pi} \int\limits_{-\infty}^{+\infty} \exp(-2x^2) dx = \dfrac{1}{\sqrt{2\pi}}.
\end{align}
Using the relation in Eq.~\eqref{eq:derivativeH}, the integral $Y_{ijk\ell}$ can be expressed in terms of $W_{ijk\ell}$, which leads directly to the following recursive formula:
\begin{align}
    \label{eq:recursiveY}
	Y_{ijk\ell} = & 2\left[\sqrt{ik} W_{(i-1)j(k-1)\ell} - \sqrt{i\ell} W_{(i-1)jk(\ell-1)} \right. \nonumber \\ & \left. -\sqrt{jk} W_{i(j-1)(k-1)\ell} + \sqrt{j\ell} W_{i(j-1)k(\ell-1)} \right].
\end{align}

\begin{figure}
    \centering
    \includegraphics[width=0.8\linewidth]{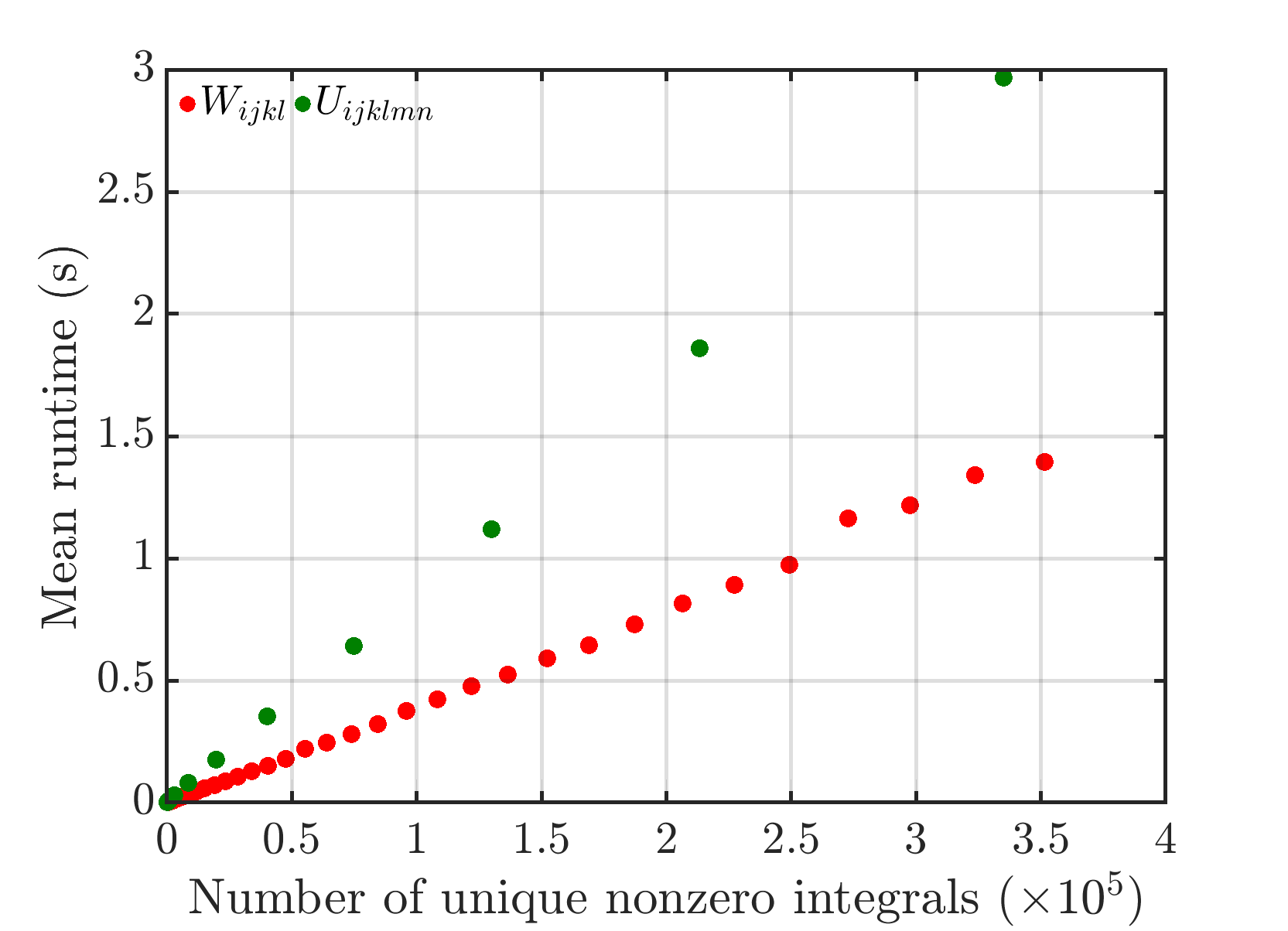}
    \caption{Mean runtime as a function of the number of unique nonzero integrals for the recursive evaluation of $W_{ijkl}$ and $U_{ijklmn}$. Each data point represents the average runtime over 10 independent runs. The standard errors of the mean are sufficiently small that the error bars are not visible on the scale of the figure. The maximum Hermite polynomial degree considered were $M=180$ for $W_{ijkl}$ and $M=50$ for $U_{ijklmn}$. The benchmark calculations were performed using Python 3.12.3 on a single CPU core of a Dell Precision 7875 Tower workstation equipped with a 16-core AMD Ryzen Threadripper PRO 7955WX CPU and 64 GB of RAM (DDR5 5600 MT/s), running Ubuntu 24.04.4 LTS.}
    \label{fig:benchmarking}
\end{figure}

\begin{table}[h!]
	\caption{\label{tab:table2} The values of the non-zero integrals $U_{ijk\ell mn}$ for the first four levels $V$. For brevity, the indices are normally sorted as $0\leq n \leq m\leq\ell\leq k\leq j\leq i$. Note that the integrals with the indices that are the permutations of the ordered set $(i,j,k,\ell,m,n)$ are identical. We use symbolic integration in Mathematica (the built-in function Integrate) to evaluate $U_{ijklmn}$ for individual sets of indices. The resulting values serve as high-precision references for validating the accuracy of the results obtained from our recursive method. Meanwhile, the Python implementation performs calculations using double-precision floating-point arithmetic. The fifth column reports the absolute difference, $\varepsilon$, between the results obtained using the Python implementation and the corresponding reference values obtained from the symbolic Mathematica calculations.}
	\begin{tabular}{|c|c|r|r|r|}
		\hline
		\textbf{Level} ${V}$               
		&   ${U_{ijklmn}}$
		&   \multicolumn{1}{c|} {\textbf{Python}}	
		&  	\textbf{Mathematica}			
		&   \textbf{$\varepsilon$} ($\times 10^{-17}$) 
		\\
		 
		\hline
		$0$
		&  $U_{000000}$                       
		&  $0.1837762984739307 $   
		&  $\frac{1}{\pi \sqrt{3}}$					
		&  $2$							
		\\
		\hline                               
		$2$
		&  $U_{200000}$   
		&  $- 0.0866329779148529 $ 
		&  $- \frac{1}{3\pi} \sqrt{\frac{2}{3}}$ 				
		&  $0$									
		\\ 
		&  $U_{110000}$
		&  $0.0612587661579769 $ 
		&  $ \frac{1}{3 \pi \sqrt{3}} $ 					
		&  $0$	 								
		\\
		
		\hline                               
		$4$
		&  $U_{400000}$  
		&  $0.0500175731198395 $
		&  $\frac{\sqrt{2}}{9\pi} $ 		
		&  $27$	 								
		\\ 
		&  $U_{310000}$
		&  $-0.0500175731198392 $  
		&  $-\frac{\sqrt{2}}{9\pi} $ 		
		&  $2$	 								
		\\ 
		&  $U_{220000}$
		&  $0.0612587661579769 $ 
		&  $\frac{1}{3 \pi \sqrt{3}} $ 				
		&  $0$	  								
		\\ 	
		&  $U_{111100}$ 
		&  $0.0612587661579769 $ 
		&  $\frac{1}{3 \pi \sqrt{3}}  $ 				
		&  $0$  								
		\\ 	
		
		\hline
		$6$     
		&  $U_{600000}$  
		&  $-0.0304397256326666 $ 
		&  $-\frac{2}{27\pi}\sqrt{\frac{5}{3}} $  	
		&  $6$  									
		\\
		&  $U_{510000}$
		&  $0.0372808978551756 $ 
		&  $\frac{\sqrt{10}}{27 \pi}  $ 	
		&  $1$								
		\\ 
		&  $U_{420000} $ 
		&  $-0.0471570201753764 $ 
		&  $-\frac{4}{27\pi} $ 	
		&  $1$								
		\\ 
		&  $U_{411000} $ 
		&  $-0.0166725243732797 $ 
		&  $- \frac{\sqrt{2}}{27 \pi}  $ 
		&  $4$	
		\\ 
		&  $U_{330000}$ 
		&  $0.0476457070117598 $ 
		&  $ \frac{7}{27 \pi \sqrt{3}} $		
		&  $2$								
		\\ 
		&  $U_{321000}$ 
		&  $0.0117892550438441 $ 
		&  $\frac{1}{27 \pi} $ 	
		&  $0$   
		\\ 
		&  $U_{311100}$ 
		&  $-0.0333450487465595 $ 
		&  $-\frac{2\sqrt{2}}{27\pi} $  	
		&  $1$								
		\\ 
		&  $U_{222000} $ 
		&  $-0.0288776593049509 $ 
		&  $-\frac{1}{9 \pi} \sqrt{\frac{2}{3}} $ 				
		&  $7$								
		\\ 
		&  $U_{221100} $
		&  $0.0204195887193256 $ 
		&  $\frac{1}{9 \pi \sqrt{3}} $ 		
		&  $3$								
		\\ 
		&  $U_{211110}  $ 
		&  $0.0288776593049509 $ 
		&  $\frac{1}{9 \pi} \sqrt{\frac{2}{3}} $ 		
		&  $7$								
		\\ 
		&  $U_{111111}  $
		&  $0.1020979435966281 $ 
		&  $\frac{5}{9 \pi \sqrt{3}} $ 		
		&  $7$								
		\\ 
        \hline 
	\end{tabular}
\end{table} 

Having the recursive formulas in Eqs.~\eqref{eq:recursiveU},~\eqref{eq:recursiveW}, and~\eqref{eq:recursiveY}, the integrals $U_{ijk\ell mn}$, $W_{ijk\ell}$, and $Y_{ijk\ell}$, which have applications in quantum physics, can be efficiently and accurately computed without numerical instabilities and overflows discussed above. To this end, we provide both a Python script and a Mathematica notebook implementing the recursive formulas via the dynamic programming technique. The routines are designed to compute either a single integral or the full set of non-zero integrals. Tables~\ref{tab:table1} and~\ref{tab:table2} compare the Python and Mathematica results for $W_{ijk\ell}$ and $U_{ijk\ell mn}$, respectively, demonstrating perfect agreement at the level of machine precision, even for extremely high-order terms such as $W_{(500)\,(500)\,(500)\,(500)}$, and $W_{(1000)\,(1000)\,(1000)\,(1000)}$ for which the existing approaches involving explicit factorial expressions fail due to numerical instabilities and overflow. In Fig.~\ref{fig:benchmarking}, we demonstrate that the mean runtime of the Python implementation, which combines the recursive formulation with the dynamic programming technique, scales linearly with the number of unique nonzero integrals for both $W_{ijkl}$ and $U_{ijkl mn}$. This behavior confirms that the computational efficiency of our proposed approach is highly competitive while avoiding the known limitations of existing analytical formula. While we do not show explicitly, the implementation of $Y_{ijk\ell}$ achieves the same order of accuracy as $W_{ijkl}$, since $Y_{ijk\ell}$ is written in terms of $W_{ijkl}$ integrals. Although the Mathematica notebook nicely yields analytical expressions, we highly recommend the Python implementation due to its superior efficiency and practical performance. 

\section{Conclusion}\label{sec13}
To summarize, we have rigorously derived the recursive formula for the integral of product of $N$ Hermite polynomials, $T_{a_1\dots a_N}$, using the known relations of Hermite polynomials and the integration by parts technique. Crucially, the recursive relation provide a robust and efficient method for numerically evaluating $T_{a_1\dots a_N}$ while avoiding the numerical instabilities stemming from large-integer factorials and quadrature methods in addition to a simple selection rule determining non-zero integrals. With these advantages, the recursive relation serves as an optimal method for the practical implementation of computer codes computing the integral $T_{a_1\dots a_N}$, achieving high-precision accuracy with minimal computational effort. As our primary motivation originates from ab initio simulations of correlated few-body quantum systems under one-dimensional harmonic confinement using the Configuration Interaction method, we demonstrate the two- and three-body contact-type interaction integrals $W_{ijk\ell}$, $Y_{ijk\ell}$, and $U_{ijk\ell mn}$ as the exemplary application of the recursive relation. We note that a recent work \cite{norris2025efficient} has employed the recursive relation $W_{ijk\ell}$ as the key aspect improving their numerical scheme with high-degree Hermite polynomials. We expect that the key findings of this work have broad relevance beyond our specific interests and will be valuable across a wide range of fields in physics and mathematics. Finally, we remark that the derivation of the recursion presented in this work is not limited to Hermite polynomials. The derivation presented here relies on three fundamental ingredients: (i) the recurrence relations, (ii) the first-order derivative relation that connect neighboring Hermite polynomials, and (iii) the integration-by-parts technique. Since analogous recurrence and derivative relations are also present in other families of classical orthogonal polynomials, including Laguerre, Legendre, and Jacobi polynomials, the recursive strategy developed in this work can be straightforwardly extended to related classes of integrals.

\section*{Data availability statement}
The authors declare that the data supporting the findings of this study are available within the article. The numerical subroutine implementing the recursive formula, together with the benchmarking data used to generate Fig.~\ref{fig:benchmarking}, is publicly available at the GitHub repository: \url{https://github.com/Tran-Duong-Anh-Tai/integral_Hermite_polynomials}.

\section*{Acknowledgements}
This work is supported by the Ministry of Education and Training of Vietnam under the grant number B2026-SPS-03. The authors thank Thomas Busch, Doerte Blume and Nathan Harshman for fruitful discussions. T.D.A.-T. gratefully acknowledges support by the Dodge Postdoctoral Researcher Fellowship at the University of Oklahoma (OU), and the Okinawa Institute of Science and Technology Graduate University (OIST).

\section*{Declarations}
The authors have no competing interests to declare that are relevant to the content of this article.





\bibliographystyle{elsarticle-num}
\bibliography{bibliography.bib}

@article{wang2009integrals,
  title={Integrals of products of Hermite functions},
  author={Wang, Wei-Min},
  journal={arXiv preprint arXiv:0901.3970},
doi={10.48550/arXiv.0901.3970},
  year={2009}
}

@article{rojo2020static,
  title={Static and dynamic properties of a few spin 1/2 interacting fermions trapped in a harmonic potential},
  author={Rojo-Franc{\`a}s, Abel and Polls, Artur and Juli{\'a}-D{\'\i}az, Bruno},
  journal={Mathematics},
  volume={8},
  number={7},
  pages={1196},
doi={10.3390/math8071196},
  year={2020},
  publisher={MDPI}
}

@article{tran2025quantum,
  title={Quantum correlations and spatial localization in trapped one-dimensional ultra-cold Bose-Bose-Bose mixtures},
  author={Anh-Tai, Tran Duong and Garcia March, Miguel Angel and Busch, Thomas and Fogarty, Thom{\'a}s},
  journal={New J. Phys.},
  year={2025},
  doi={10.1088/1367-2630/ade89e}
}

@article{titchmarsh1948some,
  title={Some integrals involving Hermite polynomials},
  author={Titchmarsh, EC},
  journal={J. London Mathematical Society},
  volume={1},
  number={1},
  pages={15--16},
  year={1948},
doi={10.1112/jlms/s1-23.2.135},
  publisher={Oxford University Press}
}

@article{norris2025efficient,
  title={Efficient Determination of Eigenenergies and Eigenstates of {N} ({N}= 3--4) Identical 1D Bosons and Fermions Under External Harmonic Confinement},
  author={Norris, J D and Blume, D},
  journal={Few-Body Systems},
  volume={67},
  number={1},
  pages={7},
  year={2026},
  publisher={Springer},
  doi={https://doi.org/10.1007/s00601-025-02025-4}
}

@phdthesis{talmi1952nuclear,
  title={Nuclear spectroscopy with harmonic oscillator wave-functions},
  author={Talmi, Igal},
  year={1952},
doi={10.3929/ethz-a-000089257},
url={https://doi.org/10.3929/ethz-a-000089257},
  school={ETH Zurich}
}

@article{moshinsky1959transformation,
  title={Transformation brackets for harmonic oscillator functions},
  author={Moshinsky, Marcos},
  journal={Nuclear Phys.},
  volume={13},
  number={1},
  pages={104--116},
  year={1959},
doi={10.1016/0029-5582(59)90143-9},
  publisher={Elsevier}
}

@article{brody1965tables,
  title={Tables of transformation brackets for nuclear shell-model calculations},
  author={Brody, Tomas A and Moshinsky, Marcos},
  journal={Universidad Nacional de Mexico, Mexico},
url={https://hdl.handle.net/2027/mdp.39015004506203},
  year={1965}
}

@book{weber2003essential,
  title={Essential Mathematical Methods for Physicists, ISE},
  author={Weber, H.J. and Arfken, G.B.},
  isbn={9780120598779},
  lccn={2003277216},
  url={https://books.google.com.vn/books?id=k046p9v-ZCgC},
  year={2004},
  publisher={Elsevier Science}
}

@book{giamarchi2003quantum,
    author = {Giamarchi, Thierry},
    title = {Quantum Physics in One Dimension},
    publisher = {Oxford University Press},
    year = {2003},
    month = {12},
    isbn = {9780198525004},
    doi = {10.1093/acprof:oso/9780198525004.001.0001},
    url = {https://doi.org/10.1093/acprof:oso/9780198525004.001.0001},
}

@article{sowinski2019one,
	title={One-dimensional mixtures of several ultracold atoms: a review},
	author={Sowinski, Tomasz and Garc{\'\i}a-March, Miguel {\'A}ngel},
	journal={Rep. Prog. Phys.},
	volume={82},
	number={10},
	pages={104401},
	year={2019},
	publisher={IOP Publishing},
	doi={10.1088/1361-6633/ab3a80}
}

@article{serwane2011deterministic,
	title={Deterministic preparation of a tunable few-fermion system},
	author={Serwane, Friedhelm and Z{\"u}rn, Gerhard and Lompe, Thomas and Ottenstein, TB and Wenz, AN and Jochim, S},
	journal={Science},
	volume={332},
	number={6027},
	pages={336--338},
	year={2011},
	publisher={American Association for the Advancement of Science},
	doi={10.1126/science.1201351}
}

@article{zurn2012fermionization,
	title={{F}ermionization of two distinguishable fermions},
	author={Z{\"u}rn, Gerhard and Serwane, Friedhelm and Lompe, T and Wenz, AN and Ries, Martin Gerhard and Bohn, Johanna Elise and Jochim, Selim},
	journal={Phys. Rev. Lett.},
	volume={108},
	number={7},
	pages={075303},
	year={2012},
	publisher={APS},
	  doi = {10.1103/PhysRevLett.108.075303}
}

@article{wenz2013few,
	title={From few to many: Observing the formation of a {F}ermi sea one atom at a time},
	author={Wenz, AN and Z{\"u}rn, G and Murmann, Simon and Brouzos, I and Lompe, T and Jochim, S},
	journal={Science},
	volume={342},
	number={6157},
	pages={457--460},
	year={2013},
	publisher={American Association for the Advancement of Science},
	doi={10.1126/science.1240516}
}

@article{garcia2013quantum,
	title={Quantum gas mixtures in different correlation regimes},
	author={Garcia-March, Miguel Angel and Busch, Thomas},
	journal={Phys. Rev. A},
	volume={87},
	number={6},
	pages={063633},
	year={2013},
	publisher={APS},
	  doi = {10.1103/PhysRevA.87.063633}
}

@article{garcia2013sharp,
	title={Sharp crossover from composite fermionization to phase separation in microscopic mixtures of ultracold bosons},
	author={Garcia-March, Miguel Angel and Juli{\'a}-D{\'\i}az, Bruno and Astrakharchik, GE and Busch, Th and Boronat, J and Polls, A},
	journal={Phys. Rev. A},
	volume={88},
	number={6},
	pages={063604},
	year={2013},
	publisher={APS},
	  doi = {10.1103/PhysRevA.88.063604}
}

@article{garcia2014distinguishability,
	title={Distinguishability, degeneracy, and correlations in three harmonically trapped bosons in one dimension},
	author={Garc{\'\i}a-March, Miguel Angel and Juli{\'a}-D{\'\i}az, Bruno and Astrakharchik, GE and Boronat, J and Polls, A},
	journal={Phys. Rev. A},
	volume={90},
	number={6},
	pages={063605},
	year={2014},
	publisher={APS},
	  doi = {10.1103/PhysRevA.90.063605}
}

@article{garcia2014quantum,
	title={Quantum correlations and spatial localization in one-dimensional ultracold bosonic mixtures},
	author={Garc{\'\i}a-March, Miguel Angel and Juli{\'a}-D{\'\i}az, Bruno and Astrakharchik, GE and Busch, Th and Boronat, J and Polls, A},
	journal={New J. Phys.},
	volume={16},
	number={10},
	pages={103004},
	year={2014},
	publisher={IOP Publishing},
	doi={10.1088/1367-2630/16/10/103004}
}

@article{olshanii1998atomic,
	title={Atomic scattering in the presence of an external confinement and a gas of impenetrable bosons},
	author={Olshanii, Maxim},
	journal={Phys. Rev. Lett.},
	volume={81},
	number={5},
	pages={938},
	year={1998},
	publisher={APS},
	  doi = {10.1103/PhysRevLett.81.938}
}

@article{murmann2015antiferromagnetic,
	title={Antiferromagnetic Heisenberg spin chain of a few cold atoms in a one-dimensional trap},
	author={Murmann, Simon and Deuretzbacher, Frank and Z{\"u}rn, Gerhard and Bjerlin, Johannes and Reimann, Stephanie M and Santos, Luis and Lompe, Thomas and Jochim, Selim},
	journal={Phys. Rev. Lett.},
	volume={115},
	number={21},
	pages={215301},
	year={2015},
	publisher={APS},
	doi = {10.1103/PhysRevLett.115.215301}
}

@article{murmann2015two,
	title={Two fermions in a double well: Exploring a fundamental building block of the Hubbard model},
	author={Murmann, Simon and Bergschneider, Andrea and Klinkhamer, Vincent M and Z{\"u}rn, Gerhard and Lompe, Thomas and Jochim, Selim},
	journal={Phys. Rev. Lett.},
	volume={114},
	number={8},
	pages={080402},
	year={2015},
	publisher={APS},
	doi = {10.1103/PhysRevLett.114.080402}
}

@article{zurn2013pairing,
	title={Pairing in few-fermion systems with attractive interactions},
	author={Z{\"u}rn, G and Wenz, AN and Murmann, Simon and Bergschneider, Andrea and Lompe, Thomas and Jochim, S},
	journal={Phys. Rev. Lett.},
	volume={111},
	number={17},
	pages={175302},
	year={2013},
	publisher={APS},
	doi = {10.1103/PhysRevLett.111.175302}
}

@article{garcia2016non,
	title={Non-equilibrium thermodynamics of harmonically trapped bosons},
	author={Garc{\'\i}a-March, Miguel {\'A}ngel and Fogarty, Thom{\'a}s and Campbell, Steve and Busch, Thomas and Paternostro, Mauro},
	journal={New J. Phys.},
	volume={18},
	number={10},
	pages={103035},
	year={2016},
	publisher={IOP Publishing},
	doi={10.1088/1367-2630/18/10/103035}
}

@article{mistakidis2023few,
  title={Few-body Bose gases in low dimensions—A laboratory for quantum dynamics},
  author={Mistakidis, SI and Volosniev, AG and Barfknecht, RE and Fogarty, T and Busch, Th and Foerster, A and Schmelcher, P and Zinner, NT},
  journal={Physics Reports},
  volume={1042},
  pages={1--108},
  year={2023},
doi={/10.1016/j.physrep.2023.10.004},
  publisher={Elsevier}
}

@article{rammelmuller2023modular,
  title={A modular implementation of an effective interaction approach for harmonically trapped fermions in 1d},
  author={Rammelm{\"u}ller, Lukas and Huber, David and Volosniev, Artem G},
  journal={SciPost Phys. Codebases},
  pages={012},
doi={10.21468/SciPostPhysCodeb.12},
  year={2023}
}

@article{garcia2018relaxation,
	title={Relaxation, chaos, and thermalization in a three-mode model of a Bose--Einstein condensate},
	author={Garcia-March, MA and van Frank, S and Bonneau, M and Schmiedmayer, J and Lewenstein, M and Santos, Lea F},
	journal={New J. Phys.},
	volume={20},
	number={11},
	pages={113039},
	year={2018},
	publisher={IOP Publishing},
	doi={10.1088/1367-2630/aaed68}
}

@article{anhtai2023quantum,
  title={Quantum chaos in interacting Bose-Bose mixtures},
  author={Anh-Tai, Tran Duong and Mikkelsen, Mathias and Busch, Thomas and Fogarty, Thom{\'a}s},
  journal={SciPost Physics},
  volume={15},
  number={2},
  pages={048},
  year={2023}, 
  doi={doi: 10.21468/SciPostPhys.15.2.048},
}

@article{bergschneider2019experimental,
  title={Experimental characterization of two-particle entanglement through position and momentum correlations},
  author={Bergschneider, Andrea and Klinkhamer, Vincent M and Becher, Jan Hendrik and Klemt, Ralf and Palm, Lukas and Z{\"u}rn, Gerhard and Jochim, Selim and Preiss, Philipp M},
  journal={Nature Phys.},
  volume={15},
  number={7},
  pages={640--644},
  year={2019},
doi={10.1038/s41567-019-0508-6},
  publisher={Nature Publishing Group UK London}
}

@article{harshman2020anyons,
  title={Anyons from three-body hard-core interactions in one dimension},
  author={Harshman, N L and Knapp, Adam C},
  journal={Annals of Physics},
  volume={412},
  pages={168003},
doi={10.1016/j.aop.2019.168003},
  year={2020},
  publisher={Elsevier}
}

@article{kanjilal2004nondivergent,
  title={Nondivergent pseudopotential treatment of spin-polarized fermions under one-and three-dimensional harmonic confinement},
  author={Kanjilal, Krittika and Blume, Doerte},
  journal={Phys. Rev. A},
  volume={70},
  number={4},
  pages={042709},
doi={10.1103/PhysRevA.70.042709},
  year={2004},
  publisher={APS}
}

@article{Tai2024inprepPRX,
  title = {Engineering impurity Bell states through coupling with a quantum bath},
  author = {Anh-Tai, Tran Duong and Fogarty, Thom\'as and de Mar\'{\i}a-Garc\'{\i}a, Sergi and Busch, Thomas and Garc\'{\i}a-March, Miguel A.},
  journal = {Phys. Rev. Res.},
  volume = {6},
  issue = {4},
  pages = {043042},
  numpages = {10},
  year = {2024},
  month = {Oct},
  publisher = {American Physical Society},
  doi = {10.1103/PhysRevResearch.6.043042},
}

@article{rojo2024few,
  title={Few particles with an impurity in a one-dimensional harmonic trap},
  author={Rojo-Franc{\`a}s, A and Isaule, F and Juli{\'a}-D{\'\i}az, B},
  journal={Physica Scripta},
  volume={99},
  number={4},
  pages={045408},
doi={10.1088/1402-4896/ad3301},
  year={2024},
  publisher={IOP Publishing}
}







\end{document}